\documentclass[reprint,showpacs,preprintnumbers,amsmath,amssymb,prc,floatfix]{revtex4-1}
\usepackage{color}
\usepackage{graphicx}% Include figure files
\usepackage{dcolumn}% Align table columns on decimal point
\usepackage{bm}% bold math
\usepackage[colorlinks,citecolor=blue,linkcolor=red,anchorcolor=blue,filecolor=blue,urlcolor=blue]{hyperref}

\begin{document}

\title{Neck configuration of Cm and Cf nuclei in the fission state within relativistic mean field 
formalism}
\author{M. Bhuyan$^1$}
\email{Email: bhuyan@ita.br}
\author{B. V. Carlson$^1$}
\email{Email: brettvc@gmail.com}
\author{S. K. Patra$^3$}
\email{Email: patra@iopb.res.in}
\author{Raj K. Gupta$^{4}$}
\email{Email: rajkgupta.chd@gmail.com}
\affiliation{$^1$Instituto Tecnol\'ogico de Aeron\'autica, 12.228-900 S\~ao Jos\'e dos Campos, 
S\~ao Paulo, Brazil} 
\affiliation{$^3$Institute of Physics, Sachivalaya Marg, Sainik School, Bhubaneswar 751005, India}
\affiliation{$^4$Department of Physics, Panjab University, Chandigarh, Punjab 160014, India}

\date{\today}

\begin{abstract}
A correlation is established between the neutron multiplicity and the neutrons number in the 
fission state of Curium and Californium isotopes within a microscopic study using relativistic 
mean field formalism. The study includes the isotopes of Cm and Cf nuclei near the valley of 
stability, and hence is likely to play an important role in the artificial synthesis of superheavy 
nuclei. The static fission path, the neutron$-$proton asymmetry, the evolution of the neck and 
their composition in terms of nucleon numbers are also estimated. We find a maximum ratio for 
average neutron to proton density, which is about $1.6$ in the breakdown of the liquid$-$drop 
picture for $^{248}$Cm and $^{252}$Cf. A strong dependence of the neutron$-$proton asymmetry on 
the neutron multiplicity in an isotopic chain is also observed.
\end{abstract}


\maketitle

\section{Introduction}
The first interpretation of nuclear fission was made about eight decades ago, though many features 
of this process are still in the rudimentary stage. The discovery of nuclear fission \cite{hanh39} 
was recognized as an evolution of the nuclear shape from a single compound nucleus split into two 
receding fragments \cite{meit39,bohr39}. This conceptual framework within the macroscopic-microscopic 
approach to the calculation of nuclear binding energies, provides a powerful theoretical tool for 
studies of low-energy fission dynamics. Further analysis from the microscopic theories to 
exploration of its dynamics are also prime objective at present in nuclear physics. In order to 
explain the fission properties of superheavy nuclei, it is essential to measure the shape (i.e. 
height and width) of the barriers and shape degrees of freedom 
\cite{bohr39,hof00,oga07,oga10,moll09,liu11}. In early days, the fission shapes were investigated 
by minimizing the sum of the Coulomb and surface energies using a development of the radius in the 
Liquid Drop Model (LDM). Recently, fusion studies have shown that the effects of the nuclear 
forces in the neck region (i.e. the gap between two fragments) of the deformed valley are indeed 
needed for optimizing the proximity energy of the fission process. The goal is more or less reached 
by following the studies from macroscopic-microscopic (mic-mac) model 
\cite{moll01,dobr07,iva09,kowa10,roye12,zhong14}, the extended Thomas-Fermi with Strutinsky 
integral (ETFSI) method \cite{mamd98,ghys15}, non-relativistic Skyrme-Hartree-Fock 
\cite{bur04,samy07,mina09,pei09,godd15,zhu16}, Gogny force \cite{egi00,ward05,ward12,ber17}, 
and relativistic mean field models \cite{bend03,lu06,sat10,bhu11,lu12,prasa13,lu14,bhu15,schu16}. 
%%%%%%%%%%%

The use of the adiabatic approximation in fission process is an interpretation of the potential 
energy surface (PES), an analogue of the classical phase space of Lagrangian and Hamiltonian 
mechanics. The fission point of a nucleus can be determined from the the total nuclear potential 
energy as a function of the shape coordinates relative to the ground state of the most favorable 
saddle point where the configuration evolves from a single nucleus into two separated fragments. 
The current way to deal with the splitting fragments depends on the most relevant collective 
variables of the nuclear shape, such as elongation, reflection asymmetry and neck structure that 
can be described by the multi-polarity deformations \cite{ward02,ward05,lu14,bhu15}. Furthermore 
a critical feature of the fission process is the multiplicity of neutron and/or small N=Z nuclei 
from the two fragments at the post scission point after they are accelerated by the mutual Coulomb 
repulsion \cite{mad85,sama95,sat08,sat10}. In this process, the neck is believed to be neutron 
rich and favorable for neutron emission than that of the  proton and/or $\alpha$-particle emission. 
At present, it is not possible to ascertain the true composition of the neck experimentally, which 
has the potential to reveal many important aspects of the fission dynamics. 
%%%%%%%%%%%%%%

The PES spanned by the relevant degrees-of-freedom of a fissile nucleus can be used to reveal a 
static fission path, fission lifetime, mass of the fragments and also many features of fission 
dynamics \cite{stas99,berg00,ward02,ward12a,ber17,sat10,lu14,bhu15,schu16}. To generate the neck 
structure of actinide nuclei and to determine the constituents of the neck (i.e. the average 
neutron-proton asymmetry and the neutron multiplicity) quantitatively, can be used to benchmark 
the predictive power of theoretical models \cite{mad85,koep88,fink89,sat10,bhu11,bhu15,lu14,ber17}. 
Such a study would be a step forward in the understanding of the fission dynamics of actinide 
nuclei \cite{sat10,lu14,schu16} and the synthesis process in the experimental laboratories 
available at present or/and under construction around the world 
\cite{leino95,gross00,sun03,wink08,sakurai08,muller91,geissel92,rodin03,thoe10}. Further the 
composition of the neck in the fission state of actinide nuclei may involve information regarding 
the formation of the elements in the rapid neutron capture process (i.e. $r$-process) of nuclear 
synthesis in stellar evolution \cite{gori11,koro12,just15}. In the present study we examine the 
properties of the fission state of actinides using the axially deformed relativistic mean field 
(RMF) model. 

This paper is organized as follows: In Sec. II we outline our scheme of the calculations using 
the relativistic mean field approach. The calculations and results are given in Sec. III. Finally, 
a summary and brief conclusion are given in Sec. IV.
%%%%%%%%%%%%%%%%%%%%%%%%%%

\section{Theoretical formalisms}
The microscopic self-consistent mean field calculation is one of the standard tools to investigate 
the properties of infinite nuclear matter and nuclear structure phenomena 
\cite{bur04,samy07,stas99,godd15,zhu16,berg00,ward12a,ber17,lu14,bhu15,schu16,bogu77,sero86,ring86}. 
The relativistic mean field (RMF) approach is one of the most popular and widely used formalisms 
among them. It starts with a basic Lagrangian that describes nucleons as Dirac spinors interacting 
through different meson fields. The relativistic mean field Lagrangian density, which has several 
modifications to account for various limitations of Walecka Lagrangian \cite{bogu77,sero86} for a 
nucleon-meson many body system \cite{bogu77,sero86,ring86,lala99c,bhu09,rein89,ring96,vret05,meng06,niks11,logo12,zhao12,lala09,ring96a,niko92,bur02,fuch95,niks02,bro92,ring90,lala97}, is
\begin{eqnarray}
{\cal L}&=&\overline{\psi}\{i\gamma^{\mu}\partial_{\mu}-M\}\psi +{\frac12}\partial^{\mu}\sigma
\partial_{\mu}\sigma \nonumber \\ 
&& -{\frac12}m_{\sigma}^{2}\sigma^{2}-{\frac13}g_{2}\sigma^{3} -{\frac14}g_{3}\sigma^{4}
-g_{s}\overline{\psi}\psi\sigma \nonumber \\ 
&& -{\frac14}\Omega^{\mu\nu}\Omega_{\mu\nu}+{\frac12}m_{w}^{2}\omega^{\mu}\omega_{\mu}
-g_{w}\overline\psi\gamma^{\mu}\psi\omega_{\mu} \nonumber \\
&&-{\frac14}\vec{B}^{\mu\nu}.\vec{B}_{\mu\nu}+\frac{1}{2}m_{\rho}^2\vec{\rho}^{\mu}.\vec{\rho}_{\mu} 
-g_{\rho}\overline{\psi}\gamma^{\mu}\vec{\tau}\psi\cdot\vec{\rho}^{\mu} \nonumber \\
&&-{\frac14}F^{\mu\nu}F_{\mu\nu}-e\overline{\psi} \gamma^{\mu}
\frac{\left(1-\tau_{3}\right)}{2}\psi A_{\mu}. 
\label{lag}
\end{eqnarray}
The $\psi$ is the Dirac spinor for the nucleon whose third component of isospin is denoted by 
$\tau_{3}$. Here $g_{\sigma}$, $g_{\omega}$, $g_{\rho}$ and $\frac{e^2}{4\pi}$ are the coupling
constants for the $\sigma-$, $\omega-$, $\rho-$ meson and photon, respectively. The constant 
$g_2$ and $g_3$ are for the self-interacting non-linear $\sigma-$meson field. The masses of the 
$\sigma-$, $\omega-$, $\rho-$ mesons and nucleons are $m_{\sigma}$, $m_{\omega}$, $m_{\rho}$, 
and $M$ respectively. The quantity $A_{\mu}$ stands for the electromagnetic field. The vector 
field tensors for the $\omega^{\mu}$, $\vec{\rho}_{\mu}$ and photon are given by,  
\begin{eqnarray}
F^{\mu\nu} = \partial_{\mu} A_{\nu} - \partial_{\nu} A_{\mu}  \\
\Omega_{\mu\nu} = \partial_{\mu} \omega_{\nu} - \partial_{\nu} \omega_{\mu} 
\end{eqnarray} 
and 
\begin{eqnarray}
\vec{B}^{\mu\nu} = \partial_{\mu} \vec{\rho}_{\nu} - \partial_{\nu} \vec{\rho}_{\mu}, 
\end{eqnarray}
respectively. From the above Lagrangian, we obtain the field equations for the nucleons and 
mesons. These equations are solved by expanding the upper and lower components of the Dirac 
spinors and the boson fields in an axially deformed harmonic oscillator basis, with an initial 
deformation $\beta_{0}$. The set of coupled equations are solved numerically by a self-consistent 
iteration method \cite{horo81,boguta81,price87,fink89}. The center-of-mass motion energy 
correction is estimated by the harmonic oscillator formula $E_{c.m.}=\frac{3}{4}(41A^{-1/3})$. 
The quadrupole deformation parameter $\beta_2$ is evaluated from the resulting proton and neutron 
quadrupole moments, as
\begin{eqnarray}
Q=Q_n+Q_p=\sqrt{\frac{16\pi}5} (\frac3{4\pi} AR^2\beta_2). 
\end{eqnarray}
The root mean square (rms) matter radius is defined as
\begin{eqnarray}
\langle r_m^2\rangle=\frac{1}{A}\int\rho(r_{\perp},z) r^2d\tau, 
\end{eqnarray}
where $A$ is the mass number, and $\rho(r_{\perp},z)$ is the axially deformed density. We obtain 
the potentials, nucleon densities, single-particle energy levels, nuclear radii, quadrupole 
deformations and the binding energies for a given nucleus. Converged ground state along with 
various constraint solutions can be obtained at different deformations including fission state 
of a nucleus (see the potential energy surface). 

To deal with the nuclear bulk properties of open-shell nuclei, one has to consider the pairing 
correlations \cite{karat10}. There are various methods such as the BCS approach, the Bogoliubov 
transformation and the particle number conserving methods that have been developed to treat the 
pairing effects in the study of nuclear properties including fission barriers 
\cite{zeng83,moli97,hao12}. The Bogoliubov transformation is widely used method to take pairing 
correlation into account for the drip-line region \cite{vret05,ring96a,meng06,lala99a}. In the 
case of nuclei not too far from the $\beta$-stability line, the constant gap BCS pairing approach 
provides a reasonably good description of pairing \cite{doba84}. The present analysis is based on 
the superheavy mass nuclei around the $\beta-$stability line, hence the relativistic mean field 
results with BCS treatment should be applicable. Further, to avoid difficulties in the calculations, 
we have employed the constant gap BCS approach to deal with the present mass region 
\cite{mad81,moll88,bhu09,bhu15,bhu18}.
%%%%%%%%%%%%%%%%%%%%%%%%%%%%%%%%%%%%%%%%%%%%%%%%%%%%%%%%%%%%%%%%%%%%%%%%%%%%%%%%%%%%%
%%%%%%%%%%%%%%%%%%%%%%%%%%%%%%%%%%%%%%%%%%%%%%%%%%%%%%
\begin{table*}
\caption{The RMF (NL3$^*$) results for the binding energy (BE), root-mean-square charge radii 
$r_{ch}$ and the quadrupole deformation parameter $\beta_2$ for $^{242,244,246,248}$Cm and 
$^{248,250,252,254}$Cf nuclei. The ground state, the constraint minima for first, second and 
fission states are given in the 1$^{st}$, 2$^{nd}$, 3$^{rd}$, and 4$^{th}$ row for each nucleus. 
The Finite-Range-Droplet-Model \cite{moll95,moll97}, Hartree-Fock + BCS \cite{gori01} and the 
experimental data \cite{audi13,angeli13,raman01} for the ground state configurations are given 
for comparison, wherever available. The energies are in $MeV$ and radii in $fm$.}
\renewcommand{\tabcolsep}{0.12cm}
\renewcommand{\arraystretch}{1.45}
\begin{tabular}{cccccccccccccccccc}
\hline \hline
Nucleus & \multicolumn{3}{c}{Binding Energy} & \multicolumn{3}{c}{Charge Radius} 
& \multicolumn{3}{c}{Quadrupole Deformation} \\
\hline
& RMF & Expt. \cite{audi13} & FRDM \cite{moll95} & RMF & Expt. \cite{angeli13} & HFBCS \cite{gori01} 
& RMF & Expt. \cite{raman01} & FRDM \cite{moll97} & HFBCS \cite{gori01} \\
\hline
$^{242}$Cm & 1823.92 & 1823.3 & 1823.05 & 5.933 & 5.8285& 5.90 & 0.287 & $--$      & 0.224 & 0.25\\ 
           & 1822.82 &        &         & 6.560 &       &      & 0.969 &           & 	& \\
           & 1822.51 &        &         & 8.143 &       &      & 2.313 &           & 	& \\
           & 1693.64 &        &         & 11.089&       &      & 5.036 &           & 	& \\
$^{244}$Cm & 1836.24 & 1835.8 & 1835.79 & 5.946 & 5.8429& 5.91 & 0.293 & 0.2972(17)& 0.234 & 0.25 \\
           & 1835.12 &        &		& 6.554 &       &      & 0.959 &           &	& \\
           & 1821.33 &        &		& 8.455 &       &      & 2.475 &	   & 	& \\
           & 1704.21 &        &		& 11.086&       &      & 5.010 &	   & 	& \\
$^{246}$Cm & 1847.34 & 1847.8 &	1847.86	& 5.947 & 5.8475& 5.93 & 0.293 & 0.2983(19)& 0.234 & 0.27 \\
           & 1845.75 &        &		& 6.553 &       &      & 0.921 &	   &    & \\
           & 1833.16 &        &		& 8.449 &       &      & 2.464 &	   &    & \\
           & 1714.82 &        &		& 10.982&       &      & 4.984 &	   &    & \\
$^{248}$Cm & 1860.63 & 1859.2 & 1859.28 & 5.959 & 5.8562& 5.94 & 0.290 & 0.2972(19)& 0.235 & 0.28 \\
           & 1859.31 &        &		& 6.556 &       &      & 0.916 &	   &    & \\
           & 1844.72 &        &		& 8.474 &       &      & 2.453 &	   &    & \\
           & 1724.72 &        &		& 10.965&       &      & 4.957 &	   &    & \\
$^{248}$Cf & 1861.11 & 1857.8 &	1857.82	& 5.990 & $--$  & 5.95 & 0.288 & $--$ 	   & 0.235 & 0.25 \\
           & 1859.83 &        &		& 6.624 &       &      & 0.969 &	   &    & \\
           & 1847.22 &        &		& 8.554 &       &      & 2.490 &	   &    & \\
           & 1726.41 &        &		& 11.115&       &      & 4.973 &	   &    & \\
$^{250}$Cf & 1872.90 & 1870.0 & 1870.29 & 6.001 & $--$  & 5.96 & 0.285 & 0.299 (15)& 0.245 & 0.28 \\
           & 1871.81 &        &		& 6.641 &       &      & 0.967 &	   &    & \\
           & 1859.51 &        &		& 8.568 &       &      & 2.479 &	   &    & \\
           & 1736.83 &        &		& 11.076&       &      & 4.945 &	   &    & \\
$^{252}$Cf & 1883.82 & 1881.3 & 1881.32 & 6.011 & $--$  & 5.97 & 0.278 & $--$	   & 0.236 & 0.25\\
           & 1882.64 &        &		& 6.681 &       &      & 1.081 &	   &    & \\
           & 1871.61 &        &		& 8.581 &       &      & 2.461 &	   &    & \\
           & 1710.73 &        &		& 10.972&       &      & 4.884 &	   &    & \\
$^{254}$Cf & 1893.25 & 1892.2 & 1891.69 & 6.022 & $--$  & 5.97 & 0.272 & $--$	   & 0.226 & 0.24 \\
           & 1891.96 &        &		& 6.987 &       &      & 1.083 &	   &    & \\
           & 1820.45 &        &		& 8.593 &       &      & 2.460 &	   &    & \\
           & 1820.73 &        &		& 10.843&       &      & 4.838 &	   &    & \\
\hline \hline
\end{tabular}
\label{tab1}
\end{table*}
%%%%%%%%%%%%%%%%%%%%%%%%%%%%%%%%%%%%%%%
\begin{figure}
%\vspace{0.6cm}
\begin{center}
\includegraphics[width=1.0\columnwidth]{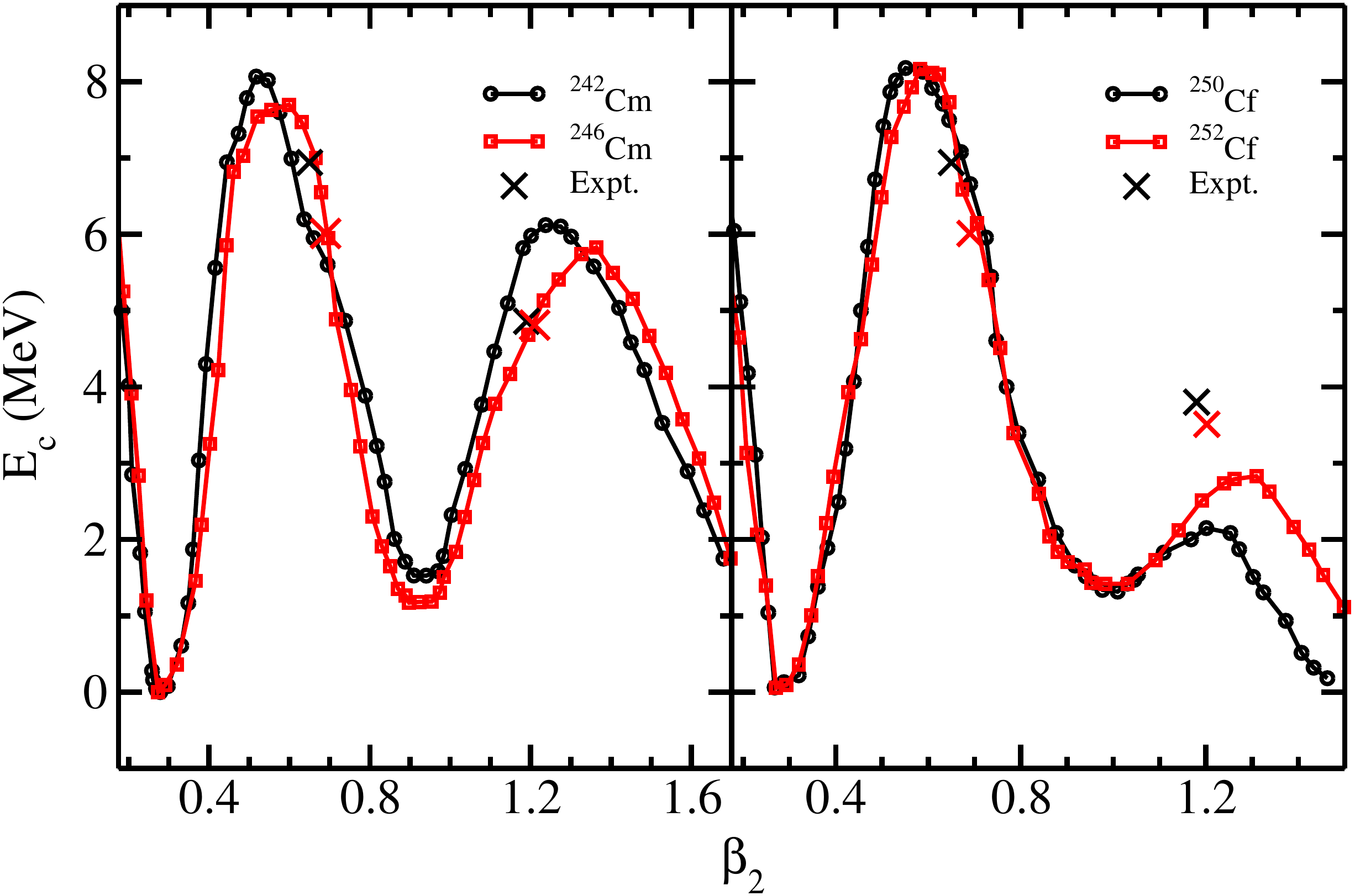}
\caption{\label{fig1} (Color online) The RMF (NL3$^*$) potential energy surfaces (PES) of 
$^{242,248}$Cm and $^{248,252}$Cf as a function of the quadrupole deformation parameter $\beta_2$ 
are displayed with the empirical values \cite{capo09} for the first and second barrier heights. Note 
that the reflection symmetry is assumed in the present calculation. Heights are in MeV. See text 
for details.}
\end{center}
\end{figure}
%%%%%%%%%%%%%%%%%%%%%%%%%%%%%%%%%%%%%%%%%%%%%%%%%%%%%%%%%%%%%%%%%%%%%%%%%%%%%%%%%%%%%%%%%%%%

\section{Calculations and Results}
In the relativistic mean field model, we performed the self-consistent calculation for maximum 
boson major shell number $N_B$ = 20 and varying maximum nucleon major shell number $N_F$ from 
14 to 24 to verify the convergence of the solutions by taking different inputs of initial 
deformation $\beta_0$ for the ground state \cite{lala97,lala09,ring90,bhu09}. From the results 
obtained, we found that the relative variations of the ground state solutions are $\leq$ 
0.004$\%$ for the binding energy and 0.002$\%$ for the nuclear radius. In the case of fission 
state solutions, the binding energy and nuclear radius varies $\leq$ 0.01$\%$ and 0.006$\%$, 
respectively over the range of major shell fermion number $N_F$ from 16 to 28 for $N_B$ = 24. 
Hence, we fixed that the number of major shells for fermions and bosons at $N_F$ = $N_B$ = 20 
and $N_F$ = $N_B$ = 24 for the ground state and for the fission state of the considered mass 
region, respectively. The number of mesh points for Gauss-Hermite and Gauss-Lagurre integral 
are $20$ and $24$, respectively. We have used the recently developed NL3$^*$ force \cite{lala09} 
for the present analysis, which is a version of the NL3 force \cite{lala97} refitted to improve 
the description for the properties of neutron- and/or proton-rich exotic and superheavy nuclei 
\cite{lala09,bhu09,bhu11}. For a given nucleus, we find various constraint solutions including 
the fission state along with the ground state (see the potential curve Fig. \ref{fig1}). The 
calculated bulk properties such as binding energy (BE), root-mean-square (rms) charge radius, 
and qudrupole deformation $\beta_2$ for the ground state, first, second, third constraint 
and fission solutions are given in the first, second, third and fourth rows for a given nucleus, 
respectively. The results obtained from NL3$^*$ force listed together with the predictions from 
Finite-Range-Droplet-Model (FRDM) \cite{moll95,moll97}, Hartree-Fock + BCS (HFBCS) \cite{gori01} 
and the experimental data \cite{audi13,angeli13,raman01}. Since BE values are not available for 
HFBCS predictions, we have listed the rms charge radius $r_{ch}$, and the quadrupole deformation 
$\beta_2$ for comparisons. We find that the ground state binding energies, charge radii and 
$\beta_2$ values agree well with the available experimental data \cite{audi13,angeli13,raman01} 
and the theoretical predictions \cite{moll95,moll97,gori01}.

As discussed above, all the isotopes of Cm and Cf are shown to have several intrinsic minima, 
where each minimum corresponds to a quadrupole deformation. For example, the ground state (g.s.), 
first excited state, second excited state and the fission state deformation $\beta_2$ for 
$^{242}$Cm are 0.287, 0.969, 2.313 and 5.036, respectively. Similarly, the values are 0.288, 
0.969, 2.490, and 4.973, respectively for $^{248}$Cf. All other isotopes and their deformations 
for various minima including the fission state are listed in Table \ref{tab1}. The solution 
corresponding to the highly deformed (hyper-deformed) configuration of $\beta_2 \sim$  2.4 for all 
isotopes provide a beautiful picture of the pre-fission state. In other words,  very smooth 
hyper-deformed solutions followed the fission configurations for all the considered isotopes in 
the present study. Further, the rms charge radius $r_{ch}$ gradually increases with increase of 
quadrupole deformation for a given nucleus.  

%%%%%%%%%%%%%%%%%%%%%
\begin{table}
\caption{The RMF (NL3$^*$) results for the first and second barrier heights of even-even isotopes 
of Cm and Cf nuclei are compared with the empirical values (Emp.) \cite{capo09}. Note that the 
reflection symmetry is assumed. Heights are in MeV.}
\renewcommand{\tabcolsep}{0.25cm}
\renewcommand{\arraystretch}{1.45}
\begin{tabular}{cccccccccccccccccc}
\hline \hline
Nucleus & \multicolumn{2}{c}{First barrier} & \multicolumn{2}{c}{Second barrier} \\
& RMF & Emp. \cite{capo09} & RMF & Emp. \cite{capo09} \\ 
\hline
$^{242}$Cm & 7.92 & 6.65 & 5.76 & 5.10 \\
$^{244}$Cm & 7.75 & 6.18 & 5.17 & 5.00 \\
$^{246}$Cm & 7.13 & 6.00 & 5.00 & 4.80 \\
$^{248}$Cm & 6.84 & 5.80 & 4.93 & 4.80 \\
$^{248}$Cf & 8.13 & $--$ & 3.33 & $--$ \\
$^{250}$Cf & 8.06 & 5.60 & 2.83 & 3.80 \\
$^{252}$Cf & 7.98 & 5.30 & 2.53 & 3.50 \\
$^{254}$Cf & 7.56 & $--$ & 1.79 & $--$ \\
\hline \hline
\end{tabular}
\label{tab2}
\end{table}

\subsection{Potential Energy Surface}
The potential energy surface (PES) is calculated by using the relativistic mean field formalism 
in a constrained procedure \cite{bhu09,bhu11,bhu15,flocard73,koepf88,kara10,lu14}, i.e., instead 
of minimizing the $H_0$, we have minimized $H'=H_0-\lambda Q_{2}$. Here, $\lambda$ is a Lagrange 
multiplier and  $Q_2$, the quadrupole moment. The term $H_0$ is the Dirac mean field Hamiltonian 
for the RMF model (the notations are standard and its form can be seen in Refs.\cite{ring90,bhu15}). 
In other words, we obtain the constrained solution from the minimization of 
$\sum_{ij}\frac{<\psi_i|H_0-\lambda Q_2|\psi_j>}{<\psi_i|\psi_j>}$ and calculate the constrained 
binding energy using $H_0$. The free energy is obtained from the minimization of 
$\sum_{ij}\frac{<\psi_i|H_0|\psi_j>}{<\psi_i|\psi_j>}$ and the converged energy solution does not 
depend on the initial guess value of the basis deformation $\beta_0$ as long as it is nearer to 
the minimum in PES. However, it converges to some other local minimum when $\beta_0$ is drastically 
different, and in this way we evaluate the different intrinsic isomeric states for a given nucleus. 
Note that the reflection symmetry is assumed for the calculation of the potential energy surface 
of the even$-$even isotopes of the Cm and Cf nuclei considered. 

The potential energy surface for $^{242,246}$Cm (left panel) and $^{250,252}$Cf (right panel) 
nuclei are shown in Fig. \ref{fig1} for a wide range of $\beta_2$ starting from the spherical 
to hyperdeformed prolate configuration. The cross (X) signs in both panels are represented by the 
empirical values \cite{capo09} of the first and second barrier heights of the respective nucleus. 
Here, we found multi-minima structure from the PES for each isotopes. In Fig. \ref{fig1}, we have 
shown the PES's of $^{242,246}$Cm and $^{250,252}$Cf as a representative case. From the figure, 
one can notice that two identical major minima exist at $\beta_2\approx$ 0.29 and 0.95 for 
$^{242}$Cm and $^{246}$Cm nuclei (see left panel of Fig. \ref{fig1}). Similarly, the minima  
also appear in case of $^{250,252}$Cf nuclei at $\beta_2\approx$ 0.28 and 0.95, respectively.
We found similar results for all the considered isotopes of Cm and Cf nuclei. The calculated 
first and second barrier heights for all the isotopes along with the empirical values 
\cite{capo09} are listed in Table \ref{tab2}. We notice that the quadrupole deformation parameters 
and the barrier heights obtained from our calculations reasonably agree with the empirical values 
\cite{angeli13,capo09} of the isotopic chains of Cm and Cf nuclei, wherever available. For 
example, the obtained first and second barrier heights for $^{242}$Cm are 7.92 and 5.76 MeV, 
respectively (see Table \ref{tab2}). Similarly, the values are 8.06 and 2.83 MeV, respectively 
for $^{250}$Cf (see Table \ref{tab2}). The corresponding empirical values for the first and second 
barrier height for $^{242}$Cm and $^{250}$Cf are 6.65, 5.10 MeV and 5.60 and 3.80 MeV, respectively. 
Moreover, the calculated mimima and/or the barriers in the PES shift a bit towards larger values 
of deformation $\beta_2$ in the isotopic chains.   

%%%%%%%%%%%%%%%%%%%%%%%%%%%%%%%%%%%%%%%%%%%%%%%%%%%%%%%%%%%%%%
\begin{figure}
%\vspace{0.6cm}
\begin{center}
\includegraphics[width=1.2\columnwidth]{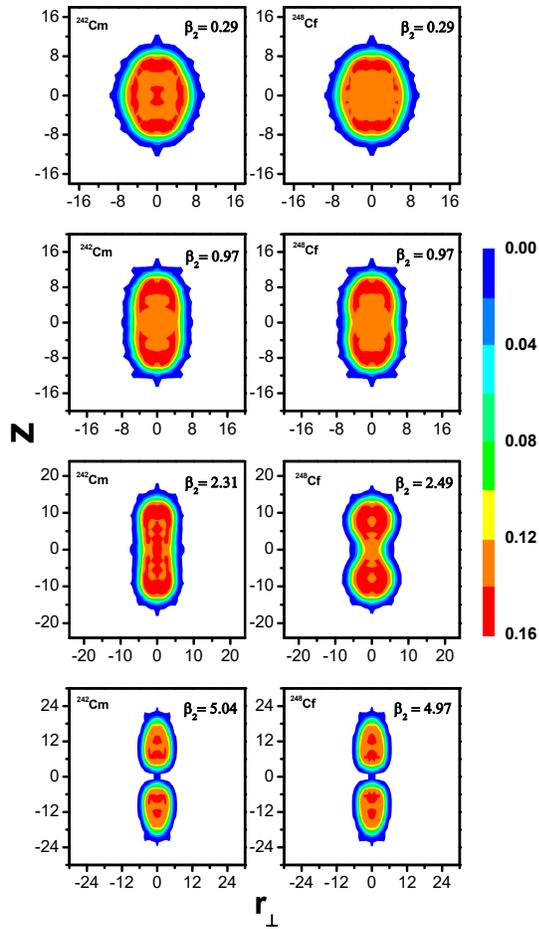}
\caption{\label{fig2}(Color online) The evolution of static fission for the isotopes of 
$^{242}$Cm (left) and $^{248}$Cf (right) for different deformations $\beta_2$ corresponding 
to the possible minima obtained in the RMF formalism using the NL3$^*$ force parameter set. 
See text for details.}
\vspace{-0.6cm}
\end{center}
\end{figure}
%%%%%%%%%%%%%%%%%%%%%%%%%%%%%%%%%%%%%%%%%%%%%%%%%%%%%%%%
\begin{figure}
\vspace{-1.0cm}
\begin{center}
\includegraphics[width=1.15\columnwidth]{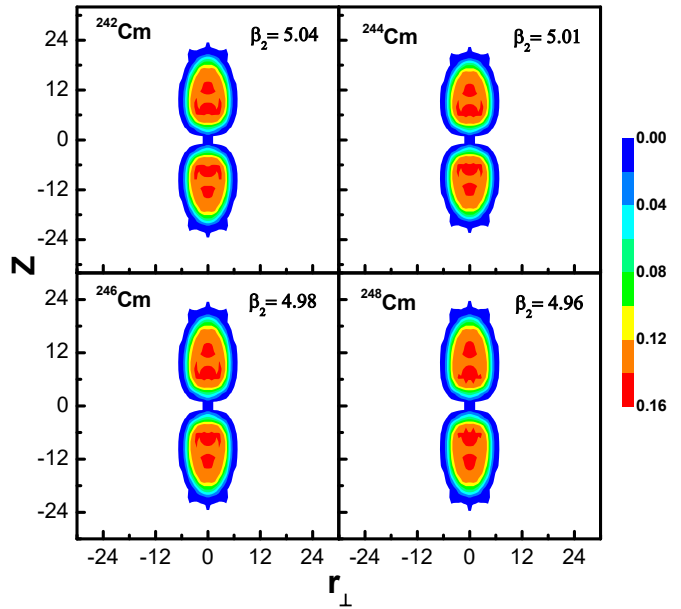}
\vspace{-2.5cm}
\caption{\label{fig3} (Color online) The RMF (NL3$^*$) total (neutron + proton) matter density 
distribution for the fission states of the $^{242,244,246,248}$Cm nuclei. See text for details.}
\end{center}
\end{figure}
%%%%%%%%%%%%%%%%%%%%%%%%%%%%%%%%%%%%%%%%%%%%%%%%%%%%%%%%%%%%%%%%%%%%%%%%%%%

\subsection{Nuclear Density Distribution}
The present calculations mainly explain the nuclear structure and sub-structure of the nucleus, 
which depend on the density distributions of the protons and neutrons for each corresponding 
state. The density distribution of the nucleus is influenced by the nuclear deformations, which 
play a prominent role in the fission study. Here, we calculate the densities for the positive 
quadrant of the plane parallel to the $z$-axis (i.e. the symmetry axis) and evaluated in the 
$zr_{\perp}$ plane, where $x^2 + y^2 = r_{\perp}^2$. The space reflection symmetries about both 
the $z$ and $r_{\perp}$ axes are conserved in our formalism. The results for the density in the 
positive quadrant can be reflected in the other quadrants to get a complete picture of the nucleus 
in the $zr_{\perp}$ plane. The unbroken space reflection symmetries of our numerical procedure 
eliminate the odd multipoles (octupole, etc.) shape degrees of freedom. In other words, there 
are limitations in explaining nuclei with an assymetric partition of particles that will not be 
properly clustered in the asymptotic limit. Nevertheless, the present study demonstrates the 
applicability of the RMF for studying the nuclear fission phenomenon and provides the scope for 
understanding the nuclear structure of even-even nuclei. Further, this furnishes an indication 
of the nuclear structure and various sub-structure for various deformed states including the 
fission state. The present calculations are performed in an axially deformed co-ordinate space. 
Consideration of the deformed coordinate space might solve some of these issues and will throw 
more light on the sub-structure of nuclei, which may be an interesting work for future. 

In Fig. \ref{fig2}, we have presented typical examples for the matter density distributions of 
the $^{242}$Cm and $^{248}$Cf nuclei for all possible solutions, starting from the ground state 
up to their static fission configuration with a neck. The shape of the $^{242}$Cm and $^{248}$Cf 
nuclei follow the deformed ground state solution around $\beta_2\approx$  0.29, and the 
super-deformed and hyper-deformed prolate solutions obtained around $\beta_2\approx$  0.97 and 
2.35, respectively. Further, a well-defined dumbbell shape of the neck configuration is reproduced 
in the RMF study as a solution of the microscopic nuclear many-body Hamiltonian around $\beta_2 
\approx$  4.50, in agreement with the age-old classical liquid drop picture of the fission process. 
The physical characteristics of the neck-structures for the isotopic chain of Cm and Cf systems 
emerging from this study will be discussed later. From Fig. \ref{fig2}, the internal configurations 
for $^{242}$Cm and $^{248}$Cf nuclei are quite evident and similar structures can found for all 
the considered isotopes of Cm and Cf. The color code, starts from deep red with maximum density 
distribution to blue bearing the minimum density. One can analyze the distribution of nucleons 
inside the various isotopes at various shapes (in black and white figures, the color code is 
read as deep black with maximum density to light gray as minimum density distribution). The 
minimum density for the oblate-state starts from 0.001 $fm^{-3}$ and goes up to a maximum of 0.16 
$fm^{-3}$ for all the shapes (see Fig. \ref{fig2}). One notices that the central density ($\rho 
\approx$ 0.16 $fm^{-3}$) becomes elongated with respect to deformation instead of changing in 
magnitudes (see the Table \ref{tab1} and Fig. \ref{fig2}). Here, we also find the neck structures 
(i.e. the elongated shape with clear-cut neck before scission) similar to those of the microscopic 
study using the constrained method with Gogny interaction \cite{dubr08} and the Skyrme-Hartree-Fock 
\cite{bonn06}. In other words, the fissioning systems energetically favor splitting into two separate 
fragments by developing an elongated shape with a neck. 

\begin{figure}
\vspace{-1.0cm}
\begin{center}
\includegraphics[width=1.13\columnwidth]{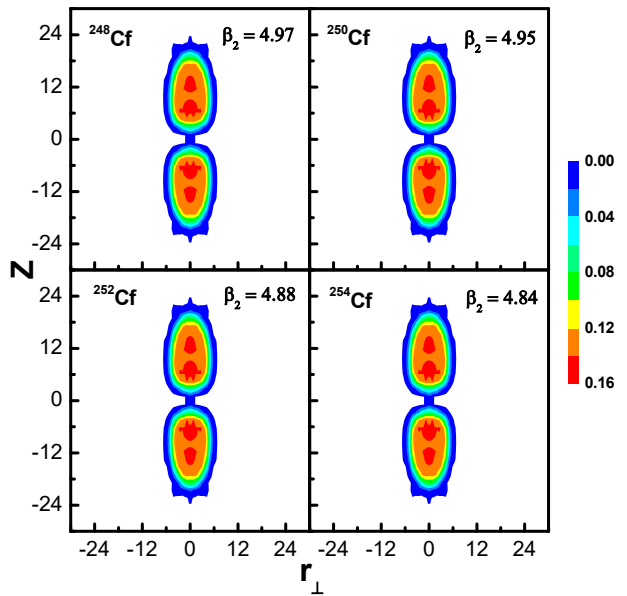}
\vspace{-2.5cm}
\caption{\label{fig4}(Color online) The RMF (NL3$^*$) total (neutron + proton) matter density 
distribution for the fission states of the $^{248,250,252,254}$Cf nuclei. See text for details.}
\end{center}
\end{figure}

Since our objective has been to critically study the neck configurations, we have presented the 
matter density distributions for the fission states of our calculations for the four isotopes of 
Cm and Cf in Figs. \ref{fig3} and \ref{fig4}, respectively. The binding energies, rms charge radii 
and quadrupole deformations of the neck configuration for $^{242,244,246,248}$Cm and 
$^{248,250,252,254}$Cf can be seen in Table \ref{tab1}. As can be seen in Fig. \ref{fig1}, the 
neck configurations lie $\approx$ 15 MeV below the respective ground states in conformity with 
the expectation and in agreement with our general notion of fission dynamics. Further, the rms 
charge radii for the neck configurations are nearly twice those of ground state, around 12 fm as 
expected. From the Figs. \ref{fig3} and \ref{fig4}, it is clear that all the isotopes undergo 
symmetric fission, which is the limitation of the present model. Here, we see how far the neck 
structure for these isotopes conform to reality from the calculated values of the first and 
second barrier height, which reasonably agree with the empirical values (see Fig. \ref{fig1} and 
Table \ref{tab2}). 

%%%%%%%%%%%%%%%%%%%%%%%%%%%%%%%%%%%%%%%%%%%%%%%%%%%%%%%%%%%%%%%%%%%%%%%%%%%%%%%%%%%%%%%%%%%%%%%%
%%%%%%%%%%%%%%%%%%%%%%%%%%%%%%%%%%%%%%%%%%%
\begin{table*}
\caption{The RMF(NL3$^*$) characteristics of neck configurations such as the quadrupole deformation 
($\beta_2$), charge radius $r_{ch}^{nk}$ of the fission state, average neutron 
($\overline{\rho}_{n}^{nk}$) and proton density ($\overline{\rho}_{p}^{nk}$) and their ratio 
($\frac{\overline{\rho}_{n}^{nk}}{\overline{\rho}_{p}^{nk}}$) in the neck region, dimension of 
the neck, length of the neck (L$^{nk}$), the number of neutron (N$^{nk}$) and proton (Z$^{nk}$ 
of the neck for $^{242,244,246,248}$Cm and $^{248,250,252,254}$Cf are presented. See text for 
details.}
\renewcommand{\tabcolsep}{0.25cm}
\renewcommand{\arraystretch}{1.45}
\begin{tabular}{ccccccccccccccccccc}
\hline \hline
Nucleus & $\beta_2$ & $r_{ch}^{nk}$ & $\overline{\rho}_{n}^{nk}$ & $\overline{\rho}_{p}^{nk}$ 
& $\frac{\overline{\rho}_{n}^{nk}}{\overline{\rho}_{p}^{nk}}$ & Range & $L^{nk}$ & $N^{nk}$ & 
$N^{nk}$ & $\frac{Z^{nk}}{N^{nk}}$ & Nucleus$^{nk}$ \\
& & & & & & ($r_1$,$r_2$; $z_1$,$z_2$) & & & \\
\hline
$^{242}$Cm& 5.036& 11.089& 0.032& 0.035& 0.91& $\pm 2.28; \pm 1.25$& 4.56& 2.01& 2.01& 1.00& $^2$He \\
$^{244}$Cm& 5.010& 11.086& 0.041& 0.034& 1.21& $\pm 2.28; \pm 1.25$& 4.56& 2.09& 2.05& 1.02& $^2$He \\
$^{246}$Cm& 4.984& 10.982& 0.047& 0.033& 1.42& $\pm 2.28; \pm 1.25$& 4.56& 2.02& 2.01& 1.01& $^4$He \\
$^{248}$Cm& 4.957& 10.965& 0.052& 0.033& 1.57& $\pm 2.28; \pm 1.25$& 4.56& 2.06& 2.01& 1.02& $^4$He \\
$^{248}$Cf& 4.973& 11.115& 0.034& 0.037& 0.92& $\pm 2.27; \pm 1.26$& 4.52& 1.01& 0.94& 1.07& $^4$H \\
$^{250}$Cf& 4.945& 11.076& 0.046& 0.036& 1.28& $\pm 2.27; \pm 1.26$& 4.52& 1.05& 0.98& 1.07& $^4$H \\
$^{252}$Cf& 4.884& 10.972& 0.051& 0.035& 1.46& $\pm 2.27; \pm 1.26$& 4.52& 2.08& 2.01& 1.03& $^4$He \\
$^{254}$Cf& 4.838& 10.843& 0.055& 0.034& 1.62& $\pm 2.27; \pm 1.26$& 4.52& 2.09& 2.01& 1.04& $^4$He \\
\hline \hline
\end{tabular}
\label{tab3}
\end{table*}
%%%%%%%%%%%%%%%%%%%%%%%%
%%%%%%%%%%%%%%%%%%%%%%%%%%%%%%%%%%%%%%%%%%%%%%%%%%%%%%%%%%%%%%%%%%%%%%%%%%%%%%%%%%%%%%%%%%%%%

\subsection{The Neck Characteristics}
The calculated yields of the total number of neutrons $N^{nk}$ and protons $Z^{nk}$ contained in 
the neck are obtained by integrating the corresponding densities over the physical dimension of 
the neck. The number of nucleons for the neck regions can calculated by, 
\begin{eqnarray}
N^{nk}=\int\int \rho_n^{nk} (r_{\perp},z) d\tau,  
\label{countN}
\end{eqnarray} 
and 
\begin{eqnarray}
Z^{nk}=\int\int \rho_p^{nk} (r_{\perp},z) d\tau,  
\label{countZ}
\end{eqnarray}
where $\rho_n^{nk}$ and $\rho_p^{nk}$ are the calculated RMF neutron and proton density distributions 
of the nucleus in the neck configuration, respectively. We also present the mean neutron and proton 
densities of the neck as, 
\begin{eqnarray}
\overline{\rho}_{n,p}^{nk} = \frac{\int \rho_{n,p}^{nk} d\tau}{\int d\tau}. 
\label{avr}
\end{eqnarray}
From Eq. \ref{avr}, we estimate the average neutron $\overline{\rho}_{n}^{nk}$ and proton 
$\overline{\rho}_{p}^{nk}$ density and their ratio $\overline{\rho}_{n}^{nk}/\overline{\rho}_{p}^{nk}$ 
for the neck region. The estimates for the neutron and/or proton constituents and their asymmetry 
are listed in Table \ref{tab3} for the $^{242,244,246,248}$Cm and $^{242,244,246,248}$Cf nuclei. As 
expected, the $\overline{\rho}_n^{nk}$ and $\overline{\rho}_p^{nk}$ for both the elements remain 
similar for all their isotopes being around 0.035 $fm^{-3}$ (see Table \ref{tab3}). The 
$\overline{\rho}_n^{nk}$ for the isotopic chains of Cm and Cf nuclei, gradually increase with the 
the neutron number. Furthermore, the neutron to proton density ratio 
$\overline{\rho}_{n}^{nk}/\overline{\rho}_{p}^{nk}$ increases gradually with respect to neutron 
number, as expected. In the isotopic chain of the Cm nuclei, the ratio has increased from 0.91 for 
$^{242}$Cm to 1.57 for $^{248}$Cm. The corresponding values are 0.92 for ${248}$Cf to 1.62 for 
$^{254}$Cf (see Table \ref{tab3}). 

We have estimated the length of the neck in the fission state, which is quite important for 
determining the neck constituents. The length of the neck $L^{nk}$ is the distance between the 
two facing connect surfaces. The width of the neck is not that important for the estimation of 
the constituents, using Eqs. \ref{countN} \& \ref{countZ}, because it only averages out the sum 
of the matter densities within $L_n$. The length of the neck $L_n$ and its constituents are 
listed in the Table \ref{tab3}. From the Table \ref{tab3}, one can find the charge radii of the 
neck configuration for all the isotopes, which are about 12 $fm$ with a well-defined neck and 
fairly extended mass distribution evident in all cases. It is indeed interesting that heavy and 
superheavy nuclei acquire such an extended dumbbell configuration, supported by the nucleon-nucleon 
force \cite{sat08,brink96}. As we move from $^{242}$Cm to $^{248}$Cm, the number of neck neutron 
and neck proton numbers remain unchanged. A similar trend is seen for the Cf isotopes. It may be 
noticed that the magnitude of the ratio $N^{nk}/Z^{nk}$ is some what different from that of the 
average neutron-to-proton neck densities $\overline{\rho}_{n}^{nk}/\overline{\rho}_{p}^{nk}$ 
(found in Table \ref{tab3}). It shows that the effective volume distributions of neutrons and 
protons are different in the neck region. The ratio of neutron-to-proton number in the neck 
region found in our present calculation is about $1.02$ for all the isotopes of Cm and Cf nuclei. 
Hence, the neck can be considered as a quasi-bound transient state of any N = Z nucleus with the 
neck nucleus correlated with those transient state being $^4$He for all isotopes of Cm nuclei. In 
the case of Cf, the effective nucleus is $np$ for $^{248,250}$Cf and $^4$He for $^{252,254}$Cf. 

\section{Summary and Conclusions}
In the present study, we have investigated the mechanism of fission decay and the shape of the 
nucleus by following the static fission path to the configuration before the breakup. The well 
established microscopic many-body nuclear Hamiltonian, i.e., the RMF theory is employed for 
estimating the classical liquid-drop picture of the fission state. The actinide isotopes of Cm 
and Cf nuclei near the valley of stability have been studied with the objective of relevance in 
stellar evolution. We found a deformed prolate configuration for the ground state of the isotopic 
chain for Cm and Cf nuclei. Furthermore, a highly deformed configuration with a neck is found by 
using a very large basis consisting of as many as 24 oscillator shells, while for the ground 
state 20 shells are adequate. This study has revealed the anatomy of the neck in the fission 
state, such as the average neutron-proton asymmetry, the length and their composition. We found 
that the average neutron-proton ratio of the neck region progressively increases with the neutron 
number in the isotopic chains of Cm and Cf nuclei. The neutron-to-proton number ratio found in 
our calculation is $1.02$, which may correlate with the quasi-bound and/or a resonance state of 
a light N = Z nucleus and /or $\alpha$-particle. The necks found in the calculation at the above 
exotic nuclei suggest a point where along with the two heavy fragments, an $\alpha-$ particle 
might be emitted at scission for the considered isotopes of Cm and Cf nuclei, except $^{248,250}$Cf. 
In case of $^{248,250}$Cf, we found the neck constitutes are to be $np$ with the two symmetry 
fragments in the fission. Due to the symmetry in the neutron-proton ratio of the neck, this cannot 
be strained into the two fragments at scission, but itself breaks down by emitting these nucleons 
which might be observed from the scission mass-yield studies. This would have strong implication 
in the energy generation of $r-$process nucleosynthesis in stellar evolution.

\section*{Acknowledgments}
This work has been supported by the FAPESP Project Nos. (2014/26195-5 \& 2017/05660-0), INCT-FNA 
Project No. 464898/2014-5, and by the CNPq - Brasil. The authors thank  Shan-Gui Zhou for his 
many-fold discussions through out the work.


\begin{thebibliography}{99}
\bibitem{hanh39}
O. Hahn and F. Strassmann, Die Naturwissenschaften {\bf 27}, 11 (1939).
\bibitem{meit39}
L. Meitner and O. R. Fritsch, Nature {\bf 143}, 239 (1939).
\bibitem{bohr39}
N. Bohr and J. A. Wheeler, Phys. Rev. {\bf 56}, 426 (1939).
\bibitem{hof00}
S. Hofmann and G. M\~unzenberg, Rev. Mod. Phys. {\bf 72}, 733 (2000).
\bibitem{oga07}
Y. Oganessian, J. Phys. G: Nucl. Phys. {\bf 34}, R165 (2007).
\bibitem{oga10}
Y. T. Oganessian {\it et al.}, Phys. Rev. Lett. {\bf 104}, 142502 (2010).
\bibitem{moll09}
P. M\~oller, A. J. Sierk, T. Ichikawa, A. Iwamoto, R. Bengtsson, H. Uhrenholt,
and S. Aberg, Phys. Rev. C {\bf 79}, 064304 (2009).
\bibitem{liu11}
Z.-H. Liu and J.-D. Bao, Phys. Rev. C {\bf 84}, 031602 (2011).
\bibitem{moll01}
P. M\~oller, D. G. Madland, A. J. Sierk, and A. Iwamoto, Nature {\bf 409}, 785 (2001).
\bibitem{dobr07}
A. Dobrowolski, K. Pomorski, and J. Bartel, Phys. Rev. C {\bf 75}, 024613 (2007).
\bibitem{iva09}
F. A. Ivanyuk and K. Pomorski, Phys. Rev. C {\bf 79}, 054327 (2009).
\bibitem{kowa10}
M. Kowal, P. Jachimowicz, and A. Sobiczewski, Phys. Rev. C {\bf 82}, 014303 (2010).
\bibitem{roye12}
G. Royer, M. Jaffre, and D. Moreau, Phys. Rev. C {\bf 86}, 044326 (2012).
\bibitem{zhong14}
C.-L. Zhong, and T.-S. Fan, Commun. Theor. Phys. {\bf 62}, 405 (2014).
\bibitem{mamd98}
A. Mamdouh, J. M. Pearson, M. Rayet, and F. Tondeur, Nucl. Phys. A {\bf 644}, 389
(1998).
\bibitem{ghys15}
L. Ghys, A. N. Andreyev, S. Antalic, M. Huyse, and P. Van Duppen, Phys. Rev. C {\bf 91},
044314 (2015).
\bibitem{bur04}
T. Burvenich, M. Bender, J. A. Maruhn, and P.-G. Reinhard, Phys. Rev. C {\bf 69},
014307 (2004).
\bibitem{samy07}
M. Samyn, S. Goriely, and J. M. Pearson, Phys. Rev. C {\bf 75}, 064312 (2007).
\bibitem{mina09}
F. Minato, S. Chiba, and K. Hagino, Nucl. Phys. A {\bf 831}, 150 (2009).
\bibitem{pei09}
J. C. Pei, W. Nazarewicz, J. A. Sheikh, and A. K. Kerman, Phys. Rev. Lett. {\bf 102},
\bibitem{godd15}
P. Goddard, P. Stevenson, and A. Rios, Phys. Rev. C {\bf 92}, 054610 (2015).
\bibitem{zhu16}
Yi Zhu, and J. C. Pei, Phys. Rev. C {\bf 94}, 024329 (2016).
\bibitem{egi00}
J. L. Egido, and L. M. Robledo, Phys. Rev. Lett. {\bf 85}, 1198 (2000).
\bibitem{ward05}
M. Warda, K. Pomorski, J. L. Egido, and L M Robledo, J. Phys. G: Nucl. Part. Phys.
{\bf 31}, S1555 (2005).
\bibitem{ward12}
M. Warda and J. L. Egido, Phys. Rev. C {\bf 86}, 014322 (2012).
\bibitem{ber17}
G. F. Bertsch, Int. J. Mod. Phys. E, {\bf 26}, 1740001 (2017)
\bibitem{bend03}
M. Bender, P.-H. Heenen, and P.-G. Reinhard, Rev. Mod. Phys. {\bf 75}, 121 (2003).
\bibitem{lu06}
H.-F. L\~u, L.-S. Geng, and J. Meng, Chin. Phys. Lett. {\bf 23}, 2940 (2006).
\bibitem{sat10}
S. K. Patra, R. K. Choudhury, and L. Satpathy, J. Phys. G: Nucl. Part. Phys. {\bf 37},
085103 (2010).
\bibitem{bhu11}
M. Bhuyan. S. K. Patra, P. Arumugam, and Raj K. Gupta, Int. J. Mod. Phys. E {\bf 20},
1227 (2011).
\bibitem{lu12}
B.-N. Lu, E.-G. Zhao, and S.-G. Zhou, Phys. Rev. C {\bf 85}, 011301(R) (2012).
\bibitem{prasa13}
V.Prassa, T.Nik\~si\'c, and D.Vretenar, Phys. rev. C {\bf 88}, 044324 (2013).
\bibitem{lu14}
B.-N Lu, J. Zhao, E.-G. Zhou, and S.-G. Zhou, Phys. Rev. C {\bf 89}, 014323 (2014).
\bibitem{bhu15}
M. Bhuyan, S. K. Patra, and Raj K. Gupta, J. Phys. G: Nucl. Part. Phys. {\bf 42},
015105 (2015).
\bibitem{schu16}
N. Schunck, and L. M. Robledo, Rep. Prog. Phys. {\bf 79}, 116301 (2016). 
\bibitem{ward02}
M. Warda, J. L. Egido, L. M. Robledo, and K. Pomorski, Phys. Rev. C {\bf 66}, 014310 (2002).
\bibitem{mad85}
P. Madler, Z. Phys. A {\bf 321}, 343 (1985).
\bibitem{sama95}
M. S. Samanta, R. P. Anand, R. K. Choudhury, S. S. Kooper and D. M. Nadkarni,
Phys. Rev. C {\bf 51}, 3127 (1995).
\bibitem{sat08}
L. Satpathy, S. K. Patra and R. K. Chouthury, PRAMANA J. Phys. {\bf 70}, 87 (2008). 
\bibitem{stas99}
A. Staszczak and Z. Łojewski, Nucl. Phys. A {\bf 657}, 134 (1999).
\bibitem{berg00}
J. F. Berger, and K. Pomorski, Phys. Rev. Lett. {\bf 85}, 30 (2000).
(2002).
\bibitem{ward12a}
M. Warda, A. Staszczak and W. Nazarewicz, Phys. Rev. C {\bf 86}, 24601 (2012).
\bibitem{koep88}
W. Koepf and P. Ring, Phys. Lett. B {\bf 212}, 397 (1988).
\bibitem{fink89}
J. Fink, V. Blum, P.-G. Reinhard, J. A. Maruhn and W. Greiner, Phys. Lett. B 
{\bf 218}, 277 (1989).
\bibitem{leino95}
M. Leino, J. \~Ayst\~o, T. Enqvist, P. Heikkinen, A. Jokinen, M. Nurmia, A. Ostrowski, 
W. H. Trzaska, J. Uusitalo, K. Eskola, P. Armbruster, and V. Ninov, Nucl. Inst. and Meth. 
Phys. Res. B {\bf 99}, 653 (1995).
\bibitem{gross00}
C. J. Gross, T. N. Ginter, D. Shapira, W. T. Milner, J. W. McConnell, A. N. James, J. W. 
Johnson, J. Mas, P. F. Mantica, R. L. Auble, J. J. Das, J. L. Blankenship, {\it et al.}, 
Nucl. Inst. and Meth. Phys. Res. A {\bf 450}, 12 (2000).
\bibitem{sun03}
Z. Sun, W. L. Zhan, Z. Y. Guo, G. Xiao, and J. X. Li, Nucl. Inst. and Meth. Phys. 
Res. A {\bf 503}, 496 (2003).
\bibitem{wink08}
M. Winkler, H. Geissel, H.Weick, B. Achenbach, K.-H. Behr, D. Boutin, A. Brünle, M. Gleim, 
W. H\~uller, C. Karagiannis, A. Kelic, B. Kindler, {\it et al.}, Nucl. Inst. and Meth. Phys. 
Res. B {\bf 266}, 4183 (2008).
\bibitem{sakurai08}
H. Sakurai, Nucl. Phys. A {\bf 805}, 526c (2008).
\bibitem{muller91}
A. C. Mueller, and R. Anne, Nucl. Inst. and Meth. Phys. Res. B {\bf 56}, 559 (1991).
\bibitem{geissel92}
H. Geissel, P. Armbruster, K. H. Behr, A. Br\~unle, K. Burkard, M. Chen, H. Folger, 
B. Franczak, H. Keller, O. Klepper, B. Langenbeck, F. Nickel, {\it et al.}, Nucl. Inst. and 
Meth. Phys. Res. B {\bf 70}, 286 (1992).
\bibitem{rodin03}
A. M. Rodin, S. V. Stepantsov, D. D. Bogdanov, M. S. Golovkov, A. S. Fomichev, S. 
I. Sidorchuk, R. S. Slepnev, R. Wolski, G. M. Ter-Akopian, Y. T. Oganessian, A. A. Yukhimchuk, 
V. V. Perevozchikov, {\it et al.}, Nucl. Inst. and Meth. Phys. Res. B {\bf 204}, 114 (2003).
\bibitem{thoe10}
M. Thoennessen, Nucl. Phys. A {\bf 834}, 688c (2010).
\bibitem{gori11}
S. Goriely, A. Bauswein, and H. T. Janka, Ast. Phys. J {\bf 738}, L32 (2011).
\bibitem{koro12}
O. Korobkin, S. Rosswog, A. Arcones, and C. Winteler, MNRAS {\bf 426}, 1940 (2012).
\bibitem{just15}
O. Just O, A. Bauswein, R. A. Pulpillo, S. Goriely and H. T. Janka, MNRAS {\bf 448}, 541 (2015).
%%%%%%%%%%%%%%%%%%%%%%%%%%
\bibitem{bogu77}
J. Boguta and A. R. Bodmer, Nucl. Phys. A {\bf 292}, 413 (1977).
\bibitem{sero86}
B. D. Serot and J. D. Walecka, in {\it Advances in Nuclear Physics}, edited by J. W.
Negele and Erich Vogt {\it Plenum Press, New York}, Vol. {\bf 16}, p. 1 (1986).
\bibitem{ring86}
W. Pannert, P. Ring, and J. Boguta, Phy. Rev. Lett., {\bf 59}, 2420, (1986).
\bibitem{lala99c}
G. A. Lalazissis, S. Raman and P. Ring, Atm. Data. Nucl. Data. Table. {\bf 71}, 1 (1999).
\bibitem{bhu09}
S. K. Patra, M. Bhuyan, M. S. Mehta and Raj K. Gupta, Phys. Rev. C {\bf 80}, 034312 (2009).
\bibitem{rein89}
P. -G. Reinhard, Rep. Prog. Phys. {\bf 52}, 439 (1989).
\bibitem{ring96}
P. Ring, Prog. Part. Nucl. Phys. {\bf 37}, 193 (1996).
\bibitem{vret05}
D. Vretenar, A. V. Afanasjev, G. A. Lalazissis, and P. Ring, Phys. Rep. {\bf 409}, 101 (2005).
\bibitem{meng06}
J. Meng, H. Toki, S. G. Zhou, S. Q. Zhang, W. H. Long, and L. S. Geng, Prog. Part. Nucl. 
Phys. {\bf 57}, 470 (2006).
\bibitem{niks11}
T. Niksi\"c, D. Vretenar, and P. Ring, Prog. Part. Nucl. Phys. {\bf 66}, 519 (2011).
\bibitem{logo12}
D. Logoteta, I. Vida\~na, C. Provid\^encia, A. Polls, and I. Bombaci, J. Phys. Conf. 
Ser. {\bf 342}, 012006 (2012).
\bibitem{zhao12}
Xian-Feng Zhao, and Huan-Yu Jia, Phys. Rev. C {\bf 85}, 065806 (2012).
\bibitem{lala09}
G. A. Lalazissis, S. Karatzikos, R. Fossion, D. Pena Arteaga, A. V. Afanasjev, P. Ring,
Phys. Lett. B {\bf 671}, 36 (2009).
\bibitem{ring96a}
P. Ring, Prog. Part. Nucl. Phys. {\bf 37}, 193 (1996).
\bibitem{niko92}
B. A. Nikolaus, T. Hoch, and D. G. Madland, Phys. Rev. C {\bf 46}, 1757 (1992).
\bibitem{bur02}
T. Burvenich, D. G. Madland, J. A. Maruhn, and P.-G. Reinhard, Phys. Rev. C {\bf 65},
044308 (2002).
\bibitem{fuch95}
C. Fuchs, H. Lenske, and H. H. Wolter, Phys. Rev. C {\bf 52}, 3043 (1995).
\bibitem{niks02}
T. Niksic, D. Vretenar, P. Finelli, and P. Ring, Phys. Rev. C {\bf 66}, 024306 (2002).
\bibitem{bro92}
R. Brockmann and H. Toki, Phys. Rev. Lett. {\bf 68}, 3408 (1992).
\bibitem{ring90}
Y. K. Gambhir, P. Ring, and A. Thimet, Ann. Phys. (N.Y.) {\bf 198}, 132 (1990).
\bibitem{lala97}
G. A. Lalazissis, J. K\"onig, and P. Ring, Phys. Rev. C {\bf 55}, 540 (1997).
\bibitem{horo81}
C. J. Horowitz and B. D. Serot, Nucl. Phys. A {\bf 368}, 503 (1981).
\bibitem{boguta81}
J. Boguta, Nucl. Phys. A {\bf 372}, 386 (1981).
\bibitem{price87}
C. E. Price, G. E. Walker, Phys. Rev. C {\bf 36}, 354 (1987).
\bibitem{karat10}
S. Karatzikos, A. V. Afanasjev, G. A. Lalazissis, P. Ring, Phys. Lett. B {\bf 689}, 72 (2010).
\bibitem{zeng83}
J. Y. Zeng and T. S. Cheng, Nucl. Phys. A {\bf 405}, 1 (1983).
\bibitem{moli97}
H. Molique and J. Dudek, Phys. Rev. C {\bf 56}, 1795 (1997).
\bibitem{hao12}
T. V. N. Hao, P. Quentin, and L. Bonneau, Phys. Rev. C {\bf 86}, 064307 (2012).
\bibitem{lala99a}
G. A. Lalazissis, D. Vretenar, P. Ring, M. Stoitsov, and L. M. Robledo, Phys. Rev. C {\bf 60},
014310 (1999).
\bibitem{doba84}
J. Dobaczewski, H. Flocard, J. Treiner, Nucl. Phys. A {\bf 422}, 103 (1984).
\bibitem{mad81}
D. G. Madland and J. R. Nix, Nucl. Phys. A {\bf 476}, 1 (1981).
\bibitem{moll88}
P. M\"oller and J.R. Nix, At. Data and Nucl. Data Tables {\bf 39}, 213 (1988).
\bibitem{bhu18}
M. Bhuyan, B. V. Carlson, S. K. Patra, and S.-G. Zhou, Phys. Rev. C {\bf 97}, 034322 (2018).
\bibitem{moll95}
P. M\"oller, J. R. Nix, W. D. Myers and W. J. Swiatecki, Atomic and Nucl. Data Tables 
{\bf 59}, 185 (1995).
\bibitem{moll97}
P. M\"oller, J. R. Nix and K. -L. Kratz, Atomic and Nucl. Data Tables {\bf 66}, 131 (1997).
\bibitem{gori01}
S. Goriely, F. Tondeur, J. M. Pearson, Atomic Data and Nucl. Data Tables {\bf 77},
311 (2001).
\bibitem{audi13}
M. Wang, G. Audi, A. H. Wapstra, F. G. Kondev, M. MacCormick, X. Xu, and B. Pfeier, Chinese
Phys. C {\bf 36}, 1603 (2013).
\bibitem{angeli13}
I. Angeli, K. P. Marinova, Atomic Data and Nucl. Data Tables {\bf 99}, 69 (2013).
\bibitem{raman01}
S. Raman, C. W. Nestor, JR., and P. Tikkanen, Atomic Data and Nucl. Data Tables {\bf 78}, 
1 (2001).
\bibitem{capo09}
R. Capote, M. Herman, P. Oblozinsky, P. Young, S. Goriely, T. Belgya, A. Ignatyuk, A. Koning, 
S. Hilaire, V. Plujko, M. Avrigeanu, O. Bersillon, M. Chadwick, T. Fukahori, Z. Ge, Y. Han, 
S. Kailas, J. Kopecky, V. Maslov, G. Reffo, M. Sin, E. Soukhovitskii, and P. Talou, {\it Special 
Issue on Nuclear Reaction Data}, Nucl. Data Sheets {\bf 110}, 3107 (2009).
\bibitem{flocard73}
H. Flocard, P. Quentin, and D. Vautherin, Phys. Lett. B {\bf 46}, 304 (1973).
\bibitem{koepf88}
W. Koepf and P. Ring, Phys. Lett. B {\bf 212}, 397 (1988).
\bibitem{kara10}
S. Karatzikos, A. V. Afanasjev, G. A. Lalazissis, P. Ring, Phys. Lett. B 689, 72 (2010).
\bibitem{dubr08}
N. Dubray, H. Goutte, and J.-P. Delaroche, Phys. Rev. C {\bf 77}, 014310 (2008).
\bibitem{bonn06}
L. Bonneau, Phys. Rev. C {\bf 74}, 014301 (2006).
\bibitem{brink96}
D. M. Brink, {\it Int. School of Physics Enrico Fermi}, Course 36 (Berlin: 
Academic), (1996).
%%%%%%%%%%%%%%%%%%%%%%%%%%%%%%%%%%%%%%%%%%%%%

\end{thebibliography}
\end{document}